\documentclass{appolb}
\usepackage[utf8]{inputenc} 
\usepackage[T1]{fontenc} 
\usepackage{graphicx}
\usepackage{amsmath}
\usepackage{url}
\usepackage{floatrow}
\UseRawInputEncoding

\newcommand{\Car}{$^{11}$C}
\newcommand{\Nit}{$^{13}$N}
\newcommand{\Oxy}{$^{15}$O}

\begin{document}
\title{$\beta^+$ radioactive nuclei created during proton therapy
\thanks{presented at 2$^{nd}$ Symposium on new trends in Nuclear and Medical Physics, 24-26 September 2025, Krak\'ow, Poland}%
}
\author{Izabela~Skwira-Chalot, Przemys{\l}aw~Sekowski, Agata~Taranienko, Adam~Spyra, Tomasz~Matulewicz \footnote{corresponding author: tomasz.matulewicz@fuw.edu.pl}  
\address{Faculty of Physics, University of Warsaw, 02-093 Warsaw, Poland} \\
\bigskip
Jan~Swako\'n
\address{Institute of Nuclear Physics PAS, 31-342 Krak\'ow, Poland} \\
\bigskip
Joanna Matulewicz
\address{National Center for Nuclear Research, Otwock, Poland}
}

\maketitle
\begin{abstract}

During proton therapy, the beam flux decreases due to inelastic interactions with nuclei.
At the highest energies used in proton therapy around 20\% protons initiate nuclear reactions. 
This report presents the cross section measurements of proton-induced production of three $\beta^+$
emitters - \Car, \Nit, \Oxy ~- with half-lives between 2 and 20 minutes, using solid C, BN and SiO$_2$ targets. 
Stacks of up to 15 targets were irradiated simultaneously with proton beams of kinetic energy below 58 MeV at the AIC-144 cyclotron of the Institute of Nuclear Physics, Polish Academy of Sciences.
The measured cross sections follow the excitation function obtained in the previous experiments, with uncertainty of a few percent. 

\end{abstract}

\section{Introduction}
Inelastic interactions with nuclei cause a reduction of proton beam flux \cite{principle,protontherapy}.
Proton beam energies are below the threshold for inelastic interactions with hydrogen, which is the dominant component of human and animal tissue.
Only three elements, in standard tissue, make up more than 1\% of the total number of atoms: oxygen, carbon and nitrogen. 
Calcium and phosphorus are present at about 0.2 \% level, 
while all other elements together contribute  less then 0.5\% \cite{composition}.
Significant fraction of protons (around 20\% at the highest energies used in proton therapy) initiates nuclear reactions.
Some of them populate $\beta^+$-decaying isotopes, that can be used to reconstruct the beam track.
This report summarizes the measurements \cite{SekowskiNIM, MatulewiczAPPB, MatulewiczEPJA} of the cross section for proton-induced production for three $\beta^+$ emitters \Car, \Nit, \Oxy ~with half lives between 2 and 20 minutes.  
Stacks of up to 15 solid (C, BN  and SiO$_2$) targets were irradiated simultaneously with the proton beam of kinetic energy below 58 MeV. 
The reactions leading to \Car, \Nit, \Oxy ~are listed in Table I.
The measured cross sections follow the excitation function obtained in  previous experiments, and have a precision of a few percent.

\vspace{3mm}
\hspace{-7.6mm} Table I: Selected nuclear reactions on carbon, nitrogen and oxygen leading to the production of \Car, \Nit, \Oxy.
     The (p,Xpn) reactions, not listed here, have Q-value lower by the deuteron binding energy of 2.3 MeV compared to (p,Xd) reactions.
\begin{table}[htb]
    \centering
    \begin{tabular}{cccc}
    \hline
    Reaction & Q-value (MeV) & Residue & $T_{1/2}$ (min)  \\
    \hline
     $^{12}$C(p,d) & -16.5 &  \\
     $^{14}$N(p,$\alpha$) & -2.9 & $^{11}$C & 20.36(2)\\
     $^{16}$O(p,$\alpha$d) & -23.7 \\
      \hline
     $^{14}$N(p,d) & -8.3 & \\
     $^{16}$O(p,$\alpha$) & -5.2 &  $^{13}$N  & 9.965(4)\\
     \hline
      $^{16}$O(p,d) & -13.4 & $^{15}$O  & 2.041(6)  \\
    \hline
    \end{tabular}
\end{table}

The $\beta^+$-decaying nuclides are particularly interesting, as the subsequent $e^+e^-$ annihilation can be detected in PET devices, enabling localization of the beam trajectory.
However, because the production of these isotopes requires relatively high Q-values (Table I), the beamÕs final path at the Bragg peak cannot be directly traced.
The use of a radioactive beam, such as \Car, has recently been the subject of intensive study \cite{radioactive1}, since the beam range can be measured.

\section{Experimental procedure}

The experimental procedure consists of two steps: irradiation and decay spectroscopy.
The activity of a sample irradiated from $t=0$ to $t=T_{EOB}$ (EOB means end of beam) can be written as
\begin{equation}
    A(t)=N_T I\sum_{C,N,O}\sigma_i\left(1-e^{-\lambda_i T_{EOB}} \right)e^{-\lambda_i t}
\end{equation}
where 
$N_T$ is the target thickness (atoms/cm$^2$),
$I$ is the beam flux,
$\sigma_i$ is the cross section for producing radioactive isotope $i$ (\Car, \Nit, \Oxy) with the decay constant $\lambda_i = \ln(2)/T_i$.
The summation runs over the reaction channels listed in the Table I.

\begin{figure}[ht]
    \floatbox[{\capbeside\thisfloatsetup{capbesideposition={right,top},capbesidewidth=5cm}}]{figure}[\FBwidth]
    {\caption{The experimental set-up. The rotating wheel (moved with a stepping motor) provides space for up to 16 irradiated targets. They are positioned, in a predefined experimental sequence, between 3 pairs of scintillation detectors. }}
     {\includegraphics[width=5cm]{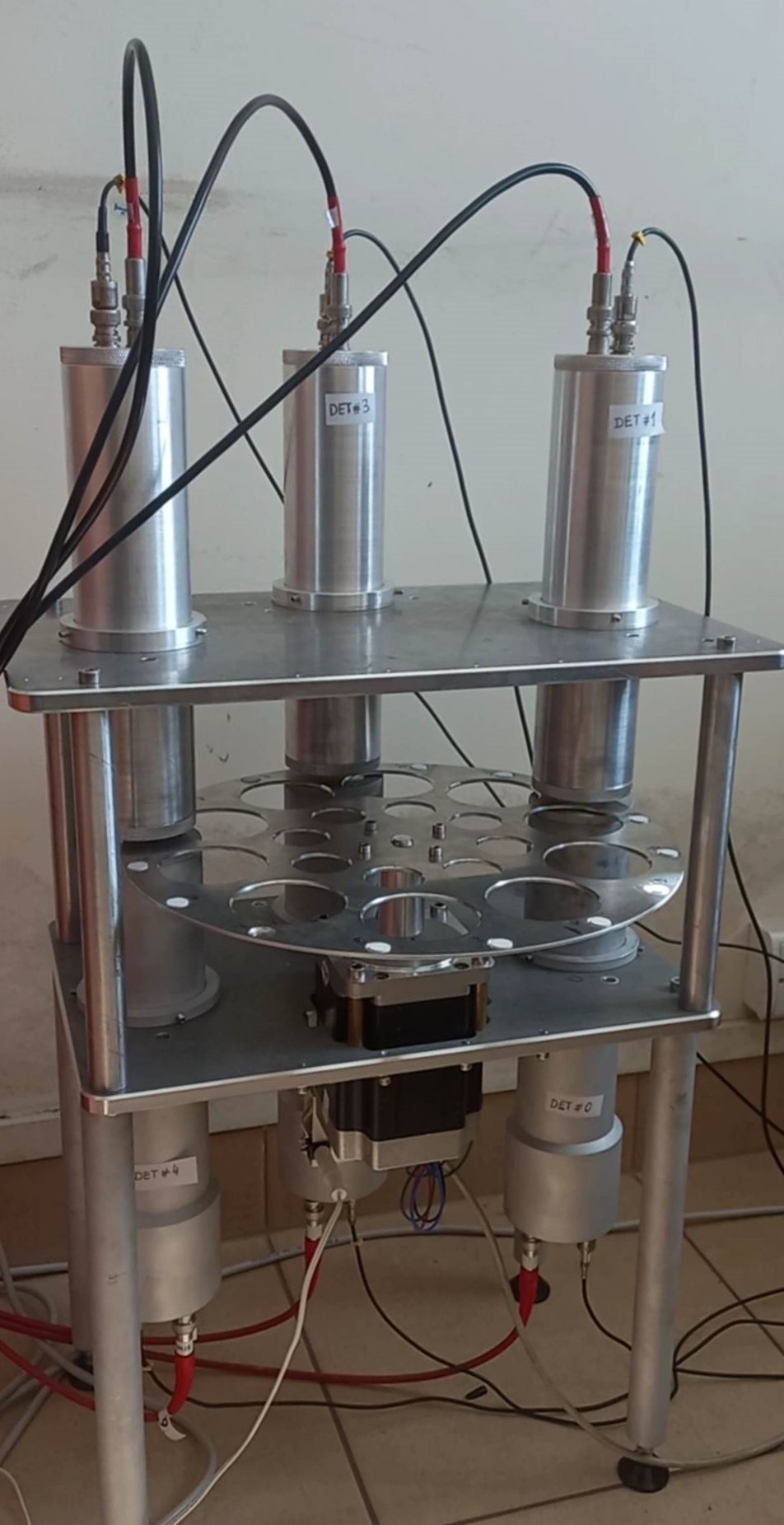}}
    \label{fig:UkladExp}
\end{figure}

In the irradiation phase a stack of several thin targets (for example, up to 16 SiO$_2$ wafers of 10 mm diameter and thickness of approximately 0.9 mm) was irradiated simultaneously with the proton beam.
This arrangement allows studying the excitation function keeping the beam flux nearly constant for all targets (the correction due to the inelastic interactions is typically below 5\% at the last target compared to the initial beam flux).
The proton energy in the center of each target was evaluated with SRIM \cite{SRIM}. 
Irradiation typically lasted a few minutes.
After irradiation, the targets were moved from the irradiation hall and placed on rotating wheel of the spectroscopy set-up \cite{SekowskiNIM}, which has spaces for up to 16 targets.
Target activities were measured with scintillation detectors.

After several upgrades, the set-up (see Fig. 1) consists of three pairs of LaBr$_3$ detectors (1 inch and 1.5 inch in each pair) readout by a CAEN digitizer (DT5730SB).
The wheel was rotated by a stepping motor in a predefined sequence, so that the activity of all irradiated targets could be recorded.
The spectroscopic measurements started typically five (or more) minutes after the irradiation ended.
This delay was caused by radiation safety procedures and manual handling of targets.
In consequence, short-lived isotopes, such as $^{14}$O with half-life of 70 s, were not measured.

The $\beta^+$ decay and the subsequent $e^+e^-$ annihilation result in the emission of two 511 keV gamma rays, registered in spectroscopic detectors.
Single detector is sensitive not only to the disintegration in the sample placed close to the detector, but also to activity from  neighboring irradiated targets \cite{Mat20}.
Lead shielding reduced this cross-talk, and a remaining neighbor contribution was corrected using the iterative procedure of Sekowski et al. \cite{SekowskiNIM}.
The triggerless readout of all detectors via CAEN DT5730SB digitizer enabled selection of coincidence events of 511 keV rays in a pair of scintillators placed opposite each other.
This mode of operation reduced the background level to a negligible level, following the observation of Horst et al. \cite{Horst2019}.
The detection efficiency in this  mode is around 1.8\%.

When the wheel rotates, activity is sampled rather then measured continuously. 
This way of operation was found to provide the same cross section (Eq. 1), as the continuous mode \cite{MatulewiczEPJA}.
The use of the same scintillation set-up for all irradiated targets reduces the systematic uncertainty in the cross section.
The measurement of the decay curve lasted typically around two hours, i.e. 6 half-lives of \Car.
Longer measurements (over 10 hours) performed on selected tissues revealed the presence of $^{18}$F with half-life of 110 min.

\section{Experimental results}

The experimental results were obtained for carbon, nitrogen, oxygen targets and for a few tissue samples.
Up to three components of $\beta^+$ activities, corresponding to the decay of \Car, \Nit ~and \Oxy, were fitted to the measured decay curves ($^{18}$F was observed in a few tissue samples). 
The presence of any other activity was not detected.
The cross section values, measured at proton beam energies below 58 MeV, have been published:
\begin{enumerate}
\item production of \Car~in $^{12}$C$(p,X)$ reactions \cite{SekowskiNIM},
\item production of \Nit~in $^{14}$N$(p,X)$ reactions \cite{MatulewiczAPPB}, and
\item production of \Car, \Nit~and~\Oxy~in $^{16}$O$(p,X)$ reactions \cite{MatulewiczEPJA}.
\end{enumerate}
The comparison of model calculations to the measured activities of irradiated tissue samples is in progress. 

\subsection{Carbon}

The $^{12}$C$(p,d)^{11}$C reaction on natural carbon targets has been extensively studied previously, so our measurements \cite{SekowskiNIM} served largely as a proof-of-concept.
The method was based on the measurement of gamma-ray spectra, where the 511 keV peak was clearly visible.
As mentioned above, the presence of the neighboring irradiated targets influences the measured activity, so the iterative correction procedure was developed, tested and applied.
The resulting measured cross section is in general agreement with numerous past experiments (except one).

\subsection{Nitrogen}

Pure nitrogen as a gaseous or liquid (cryogenic) target is not well suited to the detection method developed here, which requires solid targets.
The BN targets were selected for this research, as solid samples are manufactured by sintering the powder material for the condensed matter research.
The boron component restricted the studies only to the reaction $^{14}$N$(p,d)^{13}$N, because the $^{14}$N$(p,\alpha)^{11}$C reaction populates the same final nucleus as the $^{11}$B$(p,n)^{11}$C reaction on the abundant heavier isotope of boron.
Limited availability of appropriate BN samples allowed only a few irradiations per beam time \cite{MatulewiczAPPB}, so those decay curves were obtained in the continuous operation mode.
The measured cross sections agree well with the results obtained 60 years ago in Orsay with NaI(Tl) spectrometers by L. Valentin \cite{Valentin}, but offer improved precision.

\subsection{Oxygen}

Similarly to nitrogen, solid oxygen is only available as a compound.
The SiO$_2$ wafers have been selected \cite{apcom2023}. 
Proton-induced reactions on Si lead to the production of short-lived nuclides (T$_{1/2}$ below 10 s), which would decay before spectroscopic acquisition.
SiO$_2$ wafers of requested dimensions and optical surface quality can be ordered, making them a perfect target.
Apart of SiO$_2$ targets, previous experiments involved oxygen gas targets (up to 10 bar), BeO, MgO$_2$, water etc. \cite{MatulewiczEPJA}.
The analysis of the decay curve accounts for three produced isotopes \Car, \Nit ~and \Oxy.
Our measured production of \Oxy ~\cite{MatulewiczEPJA} closely matches the results obtained through the irradiation of bulk SiO$_2$ crystal and recording the decay of Cherenkov radiation \cite{Cherenkov}.
As relative results, Cherenkov-based measurements were normalized at one proton energy.
The observed agreement supports this choice of normalization point.
The production of \Nit ~begins at a much lower proton energy and peaks at around 15 MeV, before falling to 1-3 mb at proton energy of around 30 MeV. 
These cross sections are determined with worse precision than other channels.
In the case of \Car ~production, again the actual measurements support the results obtained with the detection of the Cherenkov radiation.
Surprisingly, the recent results of Rodriguez et al. \cite{Rodriguez} are systematically 20\% above ours and most earlier measurements.

\subsection{Tissue}

\begin{figure}[ht]
    \centering
    \includegraphics[width=8cm]{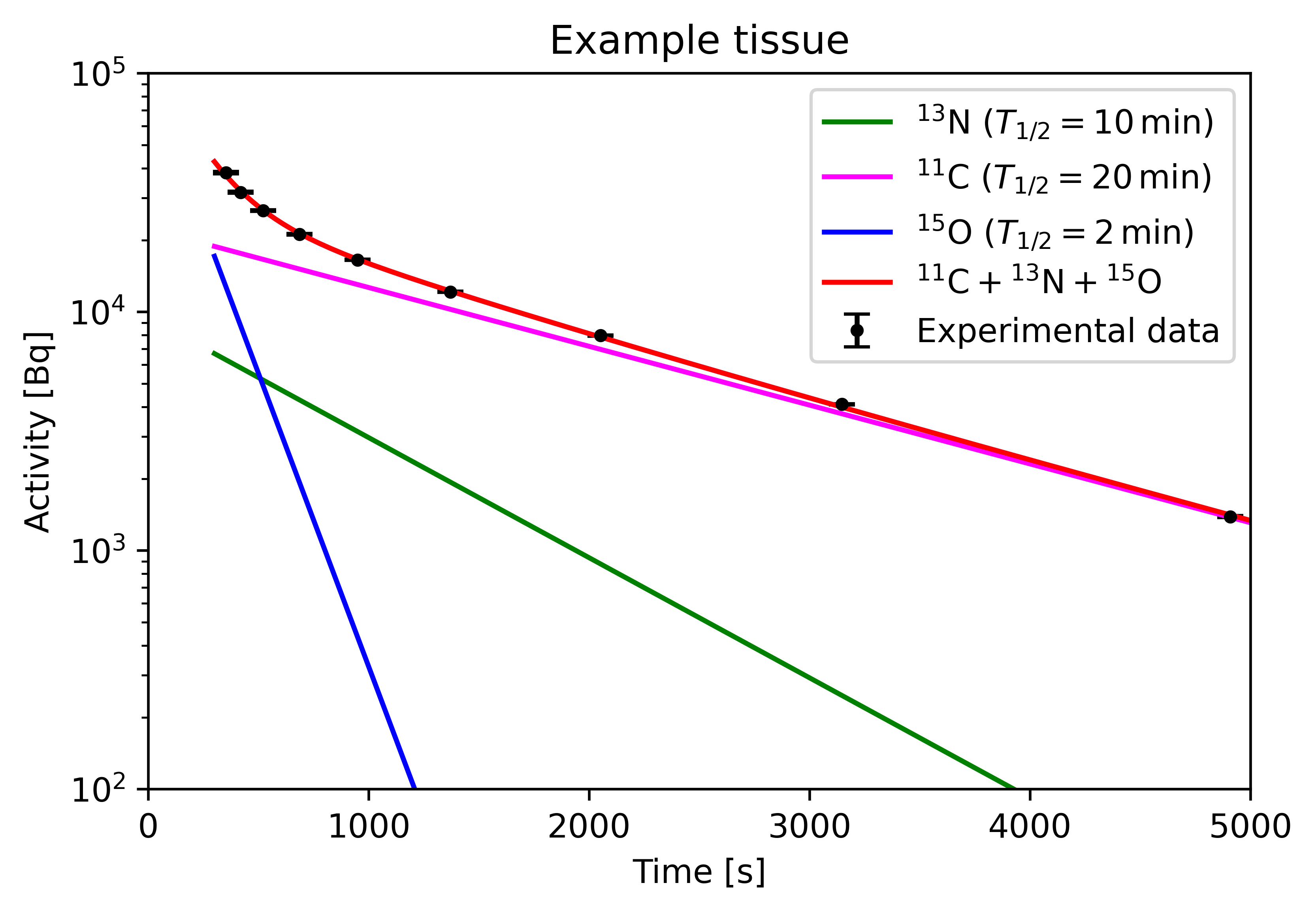}
    \caption{The measured activity of an irradiated tissue sample, interpreted as the sum of \Car, \Nit ~and \Oxy ~decays. 
    The uncertainty is statistical only.
    The time scale starts at the end of irradiation of the sample.}
    \label{fig:Decay}
\end{figure}

Several sets of tissue samples were irradiated: pig heart, pig liver, beef bone.
The 3D-printed cylindrical targets were filled with homogenized tissue and sealed on both sides with mylar foil. 
After sealing the samples were preserved at low temperature.
The experimental procedure was the same as for previously mentioned targets, in particular SiO$_2$.
The activity of each sample was interpreted as the sum of \Car, \Nit ~and \Oxy ~decays. 
Typical example is shown in Fig. \ref{fig:Decay}.

\begin{figure}[ht]
    \centering
    \includegraphics[width=8cm]{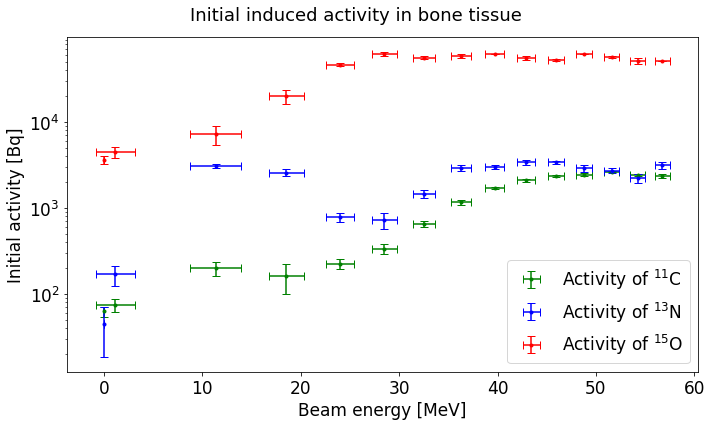}
    \caption{The activity of the bone samples irradiated by protons, measured at the end of the irradiation with 100 Gy dose (preliminary results). 
    The uncertainties come from the precision the decay curve was decomposed.}
    \label{fig:Activities}
\end{figure}

The activity at the end of irradiation of \Car, \Nit ~and \Oxy ~ isotopes was determined (preliminary results).
An example is shown in Fig. \ref{fig:Activities}.
The activity of \Oxy ~is dominant, saturating above proton energy of 30 MeV.
The notable change is visible below 20 MeV proton energy, namely the activity due to \Nit ~significantly rises.
This increase reflects the peak in the excitation function of $^{16}$O$(p,\alpha)^{13}$N reaction, while the importance of other channels tends to decrease.
The calculations accounting for the composition of the tissue are in progress.

Other $\gamma$ activities were also observed in tissue samples.
In some cases, activity corresponding to $^{18}$F was observed.
The measurements with HPGe detector placed in low-background shield revealed the presence of isotopes like $^{34m}$Cl or $^{44}$Sc \cite{Sek18}.
A copper tube, empty and filled with tissue, irradiated both using a pencil proton beam, showed enhanced production (factor $\sim$2) of isotopes such as $^{56}$Mn or  $^{61}$Cu - likely via neutron capture on stable isotopes. 
This result points to the presence of neutrons, probably from (p,pn) reactions on tissue elements.
The precise measurements with dedicated neutron detectors are needed.

\section{Conclusions}

The proton-induced reactions on nuclei abundant in tissue were performed at proton energies below 60 MeV.
The cross section of the production of $\beta^+$ decaying isotopes \Car, \Nit, and \Oxy ~were obtained.
The values are determined with uncertainties of a few percent, except in the region of single milibarns, where precision is worse.
The results follow the shape of the excitation function obtained in several (but not all) experiments performed in the past 60 years.
The experiments at higher beam energies (above 60 MeV and below 230 MeV) are planned, since that energy region remains incompletely explored. 

The cross section values, gathered during past 6 decades, vary widely in precision and are sometimes contradictory.
The importance of these nuclear physics results for the medical applications increases.
This situation calls for the critical evaluation of the nuclear physics results.
Analogous to the Particle Data Group's role in particle physics, a dedicated evaluation group should be established to provide vetted cross-section values for reactions on C, N, and O at therapy-relevant beam energies. 


\end{document}